\documentclass[twocolumn,showpacs,preprintnumbers,amsmath,amssymb]{revtex4}
\usepackage{graphicx}% Include figure files
\usepackage{bm}% bold math

\begin{document}

%\preprint{APS/123-QED}

\title{AC Josephson current and supercurrent noise through one-dimensional correlated electron systems}% Force line breaks with \\

\author{Nobuhiko Yokoshi}
\author{Susumu Kurihara}%
 %\email{yokoshi@kh.phys.waseda.ac.jp}
\affiliation{%
Department of Physics, Waseda University, Okubo, Shinjuku, Tokyo 169-8555, Japan }%

\date{\today}% It is always \today, today,

\begin{abstract}
AC Josephson effect in one-dimensional Tomonaga-Luttinger liquid (TLL) adiabatically connected to superconducting electrodes is theoretically investigated. It is found that density fluctuations due to repulsive electron-electron interactions in TLL inhibit Josephson oscillations, whereas they do not affect time-independent current part. We also show that the fluctuations reduce supercurrent noise caused by multiple Andreev reflections. This indicates that the quantum fluctuations in TLL disturb the superconducting phase coherence spreading across the junction.
\end{abstract}

\maketitle

%\section{\label{sec:level1}First-level heading:\protect\\ The line
%break was forced \lowercase{via} \textbackslash\textbackslash}

A superconducting weak link is a probable stage for inhomogeneous superconductivity. This is because superconducting phase coherence is sustained across the weak link, and should be strongly affected by various nature of intermediate segment sandwiched~\cite{Josephson}. In general, as the spatial dimension is reduced, thermal or quantum fluctuations tend to disturb the long-ranged phase correlation. Then, Josephson effect through low-dimensional system is essentially exposed to the disturbances, and is obliterated at low temperatures~\cite{Hermele}. 

Specifically one-dimensional (1D) electron systems are sensitive to inter-particle interactions. Focusing on low energy regime, they are believed to behave as Tomonaga-Luttinger liquid (TLL)~\cite{Giamarchi}. In such a state, the phase correlations are no longer infinitely long-ranged but exhibit only quasi-long-range order. Then, the correlators decay following power law~\cite{Kane}. On the other hand, a couple of experiments recently reported supercurrent flow~\cite{Kasumov1,Kasumov2,Morpurgo,Lindelof} and proximity-induced superconductivity~\cite{Haruyama} in carbon nanotubes (CNTs) suspended between superconductors. Since the metallic CNTs are ideal 1D conductors and the excitations in them can be described as TLL~\cite{Egger}, it can be said that these experiments provided eligible stages to investigate superconducting coherence in 1D correlated systems.

Theoretically, DC Josephson current through TLL has been studied for the past decade by many authors~\cite{Fazio,Maslov,Takane,Affleck,Caux,Yokoshi}. Compared with DC current, however, the study of AC Josephson current was limited in low transparency region~\cite{Fazio2}. In this work, we investigate AC Josephson effect through TLL adiabatically connected with superconducting electrodes. As for current-voltage ($I$-$V$) characteristics, it is found that the density fluctuations due to the repulsive interactions compress the Josephson oscillations, while the time-independent current is not affected. We also show that the shot noise caused by multiple Andreev reflections~\cite{Averin2} is crucially suppressed. The suppression can be explained in the framework of Caldeira-Leggett model, which describes the effect of dissipative environment on macroscopic quantum tunneling~\cite{Caldeira}. These results indicate that the low-lying excitations in TLL disturb the phase coherence across the Josephson junctions.

We suppose identical $s$-wave superconductors with energy gap $\Delta$ for the reservoirs (electrodes). The interfaces between TLL and the electrodes are modeled as the adiabatic openings of many channels so that we can simulate a bulk superconductor with its subdivision narrowed to form a wire. In the TLL region, the Coulomb interactions are assumed to be point contact type. For simplicity we neglect the processes with back-scattering and umklapp-scattering, i.e., only the electron density with long wave length is essential. Then we approximate that the interactions are switched off abruptly at the interfaces. 

Andreev reflections discussed below are performed by individual electrons in TLL and the superconductors. Then it is convenient to employ the method in which single particle excitations are treated in parallel with the low energy fluctuations. For that purpose, in the 1D region, we start with the action using auxiliary fields which incorporate the forward-scatterings~\cite{Strato};
%%%%%%%%%%%%%%%%%%%%%%%%%%%%%%%%%%%%%%%%%%%%%%%%%%%%%%%%%%%%%%%%%%%%%%%%%%%%%%
%%%%%%%%%%%%%%%%%%%%%%%%%%%%%%%%%%%%% action %%%%%%%%%%%%%%%%%%%%%%%%%%%%%%%%%
\begin{eqnarray}
\!\!\!\! S[\phi]&=&\int dtdx \Bigl[
L_0(\psi^{\dagger},\psi)+L_1(\phi) \nonumber \\
&&~~~~~~~~~~~~~~~~~~
+\sum _{a,s} \phi_{a,s}(x,t) \rho_{a,s}(x,t)
\Bigr],
\label{action}
\end{eqnarray}
%%%%%%%%%%%%%%%%%%%%%%%%%%%%%%%%%%%%%%%%%%%%%%%%%%%%%%%%%%%%%%%%%%%%%%%%%%%%%
where $\rho_{a,s}(x,t)=\psi_{a,s}^{\dagger}\psi_{a,s}$ is chiral density operator. $a=\pm$ and $s=\pm$ denote direction of movement and spin, respectively. $L_0$ and $L_1$ are the Lagrangian density of free fermions propagating with Fermi velocity $v_F$ and of the density fluctuations induced by the interactions;
%%%%%%%%%%%%%%%%%%%%%%%%%%%%%%%%%%%%%%%%%%%%%%%%%%%%%%%%%%%%%%%%%%%%%%%%%%%%%
%%%%%%%%%%%%%%%%%%%%%%%%%%%%%%%%%%  L0, Ll %%%%%%%%%%%%%%%%%%%%%%%%%%%%%%%%%%
\begin{eqnarray}
L_0&=&\sum _{a,s}\psi_{a,s}^{\dagger}(x,t)
(i\frac{\partial}{\partial t} + iav_F \frac{\partial}{\partial x})\psi_{a,s}(x,t) , \\
L_1&=&[\phi,\hat{{\rm g}}^{-1}\phi],
\end{eqnarray}
%%%%%%%%%%%%%%%%%%%%%%%%%%%%%%%%%%%%%%%%%%%%%%%%%%%%%%%%%%%%%%%%%%%%%%%%%%%%%%
with the vector $\phi=(\phi_{+\uparrow},\phi_{-\downarrow},\phi_{-\uparrow},\phi_{+\downarrow})^T$ and $\hat{{\rm g}}$ being the $(4 \times 4)$ interaction matrix. Throughout the work, we set $\hbar = k_B =1$. The auxiliary field $\phi_{a,s}(x,t)$ acts as a fluctuating electrical potential. Therefore the net quantities of the charge density and the current are obtained after taking a functional average in terms of $S_{\rm ind}[\phi]=\int dt dx L_1$. We can transform the action to the Gaussian form of chiral fields defined by $(\partial_t+av_F\partial_x)\theta_{a,s}(x,t)=\phi_{a,s}(x,t)$~\cite{Lee,Grishin}. Thus one can rewrite the problems to the ones of the free electrons propagating in integrable {\it internal environment}.

Because TLL describes only low energy physics, we treat the free fermion part with quasiclassical model to keep consistency in the approximation. In addition the voltage drop in TLL is disregarded approximately. One can thus obtain retarded (advanced) Green's functions in TLL by superposing formal solutions of following Eilenberger equation~\cite{Eilenberger}
%%%%%%%%%%%%%%%%%%%%%%%%%%%%%%%%%%%%%%%%%%%%%%%%%%%%%%%%%%%%%%%%%%%%%%%%%%%%%%
%%%%%%%%%%%%%%%%%%%%%%%%%%% Eilenberger equation %%%%%%%%%%%%%%%%%%%%%%%%%%%%%
\begin{eqnarray}
&&iv_F \frac{\partial}{\partial x}\hat{g}^{R (A)}(x,t,t'|\phi) \nonumber \\
&&+\Bigl[
i  \frac{\partial}{\partial t}
 \hat{\tau}_z \hat{\Sigma }_z 
+\hat{\phi}(x,t) \hat{\Sigma }_z, \ \hat{g}^{R (A)}(x,t,t'|\phi) 
\Bigr]_-=0,
\end{eqnarray}
%%%%%%%%%%%%%%%%%%%%%%%%%%%%%%%%%%%%%%%%%%%%%%%%%%%%%%%%%%%%%%%%%%%%%%%%%%%%%%
where $[ \cdots ]_-$ denotes a commutator as well as convolution integral in terms of the internal time, and
%%%%%%%%%%%%%%%%%%%%%%%%%%%%%%%%%%%%%%%%%%%%%%%%%%%%%%%%%%%%%%%%%%%%%%%%%%%%%%
%%%%%%%%%%%%%%%%%%%%%%%%%%%%       Sigma       %%%%%%%%%%%%%%%%%%%%%%%%%%%%%%%
\begin{eqnarray}
\hat{\tau}_i= 
\left(
\begin{array}{cc}
{\boldsymbol \sigma}_i & 0 \\ 0 & {\boldsymbol \sigma}_i
\end{array}
\right), \ \ \ \ 
\hat{\Sigma}_z= 
\left(
\begin{array}{cc}
{\bf 1} & 0 \\ 0 & {\bf -1}
\end{array}
\right)
\end{eqnarray}
%%%%%%%%%%%%%%%%%%%%%%%%%%%%%%%%%%%%%%%%%%%%%%%%%%%%%%%%%%%%%%%%%%%%%%%%%%%%%%
%%%%%%%%%%%%%%%%%%%%%%%%%%%%%%%%%%%%%%%%%%%%%%%%%%%%%%%%%%%%%%%%%%%%%%%%%%%%%%
with ${\boldsymbol \sigma}_i$s being usual Pauli matrices. Here, quantities with ``hat'' denote $(4 \times 4)$ matrices, and those with boldface $(2 \times 2)$ matrices. 1st and 3rd rows correspond to right and left moving electrons with spin up, whereas 2nd and 4th rows to left and right moving holes with spin down. In a similar fashion, the quasiclassical Green's functions in superconductors can be calculated. Here we assume that the influence of the density waves in TLL falls off in the superconductors, and neglect the charge fluctuations far from the interfaces. This is because the superconducting energy gap $\Delta$ in the spectrum prevents the gapless modes from exciting.

Since we focus on the junctions with clean interfaces, the boundary condition at $x=\pm L/2$ reduces to~\cite{Zaitsev}
%%%%%%%%%%%%%%%%%%%%%%%%%%%%%%%%%%%%%%%%%%%%%%%%%%%%%%%%%%%%%%%%%%%%%%%%%%%%%%
%%%%%%%%%%%%%%%%%%%%%%%%%% boundary condition  %%%%%%%%%%%%%%%%%%%%%%%%%%%%%%%
\begin{eqnarray}
\hat{g}^p(\pm \frac{L}{2}-0,t,t'|\phi)=\hat{g}^p(\pm \frac{L}{2}+0,t,t'|\phi),
\end{eqnarray}
%%%%%%%%%%%%%%%%%%%%%%%%%%%%%%%%%%%%%%%%%%%%%%%%%%%%%%%%%%%%%%%%%%%%%%%%%%%%%%
where $p=\{R, A, K \}$ denotes the retarded, the advanced and the Keldysh part. We choose zero of energy at Fermi level of TLL, i.e., the one of the left (right) electrode is shifted to $\pm eV/2$. A quasiparticle in TLL performs a set of back-and-forth Andreev reflections for each Cooper pair tunneling. Then, the Green's functions satisfy recurrence equations for the transferred charge~\cite{Averin,Gunsen}. One can easily find that they acquire the phase shift during each Cooper pair tunneling~\cite{Yokoshi}
%%%%%%%%%%%%%%%%%%%%%%%%%%%%%%%%%%%%%%%%%%%%%%%%%%%%%%%%%%%%%%%%%%%%%%%%%%%%%%
%%%%%%%%%%%%%%%%%%%%%%%%%%%%%%%%   \Phi    %%%%%%%%%%%%%%%%%%%%%%%%%%%%%%%%%%%
\begin{eqnarray}
\Phi_{s}(t,0)=\theta_{a,s} \bigl(\frac{L}{2},0 \bigr)+\theta_{-a,-s}\bigl(\frac{L}{2},0 \bigr) - \{ \frac{L}{2} \rightarrow -\frac{L}{2} \}, 
\label{Phi}
\end{eqnarray}
%%%%%%%%%%%%%%%%%%%%%%%%%%%%%%%%%%%%%%%%%%%%%%%%%%%%%%%%%%%%%%%%%%%%%%%%%%%%%%
which reflects the singlet superconductivity of the electrodes. This means that TLL modifies the definite phase difference $2eV$ by $\Phi_{s}$, whereas the effects of TLL disappear deep in the electrodes. 

Since the Fermi wave number in TLL is shifted by $\delta \rho_{a,s}(x,t)=\partial_x \theta_{a,s}/2\pi$~\cite{Lee,Grishin}, one properly accounts for the excess charges between the interfaces through consideration of $\Phi_{s}$. In addition, the adiabatic interfaces do not hold the charge number in TLL assuming $e^2/2C \ll \Delta$, where $C$ is the capacitance representing the long-range part of the Coulomb interactions. Then, the boundary values of $\theta$s are not fixed, i.e., the momentum unit of the density waves is small compared with $\pi/L$~\cite{Yokoshi}. This claim is in common with the different procedures in treating TLL with normal metal reservoirs~\cite{LRG1,LRG2,LRG3,LRG4,LRG5} and usual Fermi liquid between superconductors~\cite{Josephson}.

Firstly we investigate the $I$-$V$ characteristics. The net AC Josephson current is calculated by averaging
%%%%%%%%%%%%%%%%%%%%%%%%%%%%%%%%%%%%%%%%%%%%%%%%%%%%%%%%%%%%%%%%%%%%%%%%%%%%%%
%%%%%%%%%%%%%%%%%%%%%%%%%%%%%%%%    current    %%%%%%%%%%%%%%%%%%%%%%%%%%%%%%%
\begin{equation}
I(t|\phi)=\frac{e}{8\pi}{\rm Tr} 
\Bigl[
\hat{\tau}_z \hat{\Sigma}_z~\hat{g}^K (t=t'|\phi)
\Bigr] \nonumber
\end{equation}
%%%%%%%%%%%%%%%%%%%%%%%%%%%%%%%%%%%%%%%%%%%%%%%%%%%%%%%%%%%%%%%%%%%%%%%%%%%%%%
over the density fluctuations. It is expressed as a combination of harmonics with the period $T_J=\pi/eV$, i.e., $I(t)=\sum_{m=-\infty}^{\infty}I_m \exp (-2mi eVt)$~\cite{Averin,Gunsen,mendo}. The amplitude of $m$-th harmonics  ($m\geq 0$) is given by
%%%%%%%%%%%%%%%%%%%%%%%%%%%%%%%%%%%%%%%%%%%%%%%%%%%%%%%%%%%%%%%%%%%%%%%%%%%%%%%
%%%%%%%%%%%%%%%%%%%%%%%%%%%%%% I-V characteristics %%%%%%%%%%%%%%%%%%%%%%%%%%%%
\begin{widetext}
\begin{eqnarray}
I_m
&=&\frac{e}{\pi}
\Bigl[
eV \delta_{0,m}
-\Lambda^{m^2}
\int d\epsilon 
\tanh [\frac{\epsilon+eV/2}{2T}] 
\Bigl(
1-A(\epsilon+\frac{1}{2}eV)
\Bigr) 
\nonumber
\\
&&~~~~~~~~~~~~~~~~~~~~~~~
\times
\sum_{n=0}^{\infty} 
\prod_{l=1}^{m}e^{2i\frac{\epsilon+(2l+n)eV}{v_F/L}}
\prod_{l=1}^{n} A \bigl( \epsilon+(l+\frac{1}{2})eV \bigr) 
\prod_{l=1}^{2m} \gamma_R \bigl( \epsilon+(l+n+\frac{1}{2})eV \bigr)
\Bigr],
\label{IV}
\end{eqnarray}
\end{widetext}
%%%%%%%%%%%%%%%%%%%%%%%%%%%%%%%%%%%%%%%%%%%%%%%%%%%%%%%%%%%%%%%%%%%%%%%%%%%%%%%
where $A(\epsilon)=|\gamma_R(\epsilon)|^2$ is the Andreev reflection probability with $\gamma_R(\epsilon)=(\epsilon-\sqrt{(\epsilon+i0)^2-\Delta^2})/\Delta$. The effect of the interactions appears only in 
%%%%%%%%%%%%%%%%%%%%%%%%%%%%%%%%%%%%%%%%%%%%%%%%%%%%%%%%%%%%%%%%%%%%%%%%%%%%%%
%%%%%%%%%%%%%%%%%%%%%%%%%  Luttinger contribution %%%%%%%%%%%%%%%%%%%%%%%%%%%%
\begin{eqnarray}
\Lambda = 
(\frac{\pi T}
{D})^{K_{\rho}^{-1}-1}
\frac{\sinh (\frac{L}{2L_T})}
{\bigl( u_{\rho} \sinh (\frac{L}{2u_{\rho}L_T})
\bigr)^{K_{\rho}^{-1}} },
\end{eqnarray}
%%%%%%%%%%%%%%%%%%%%%%%%%%%%%%%%%%%%%%%%%%%%%%%%%%%%%%%%%%%%%%%%%%%%%%%%%%%%%%%
where $K_{\rho}, u_{\rho}$ are Luttinger parameter and velocity renormalization for the charge density fluctuations. Here $K_{\sigma}=u_{\sigma}=1$ is assumed for spin part. $D$ and $L_T=v_F/2\pi T$ are high-energy cut-off and thermal length. One can see that the repulsive interactions ($K_{\rho}<1$) inhibit the Josephson oscillations. Further the inhibition is more serious as the Josephson frequency increases. On the other hand, the renormalization does not appear in the non-oscillating current part with $m=0$; the critical current is still $2e\Delta/\pi$ at absolute zero. This indicates that the collective fluctuations act only on the Andreev phase (the argument of the Andreev reflection amplitude $\gamma_R$) as far as the scattering problem is considered. 

The renormalization reflects the algebraic decay of the singlet superconductivity phase correlation between the two interfaces~\cite{Giamarchi}. As far as the power law is concerned, Eq.~(\ref{IV}) corresponds to the extension of the previous work~\cite{Fazio2} to infinite order of the tunnel Hamiltonian. However we cannnot find the $u_{\rho}$-dependent amplitude oscillation with the length of TLL, which is caused by the spin-charge separation~\cite{Fazio2}. This is because we do not consider here the voltage drop explicitly in TLL. In studying DC effect, Maslov {\it et al.} applied an extended open boundary condition including Andreev reflections to TLL so that the fluctuating potentials cannot affect the phase difference~\cite{Maslov}. We can apply the condition to the AC effect alike, which yields no renormalization of the Josephson oscillations. It is however out of scope of the present work where we consider the 1D region is adiabatically widened at the interfaces.  

Although we have investigated the average current so far, it is well-known that current fluctuation also can be used as good indicator of the phase coherence. Averin and Imam predicted that the shot noise in Josephson junctions is enhanced by the multiple Andreev reflections~\cite{Averin2}, which was verified experimentally, e.g., in atomic point contact~\cite{Cron} and superconductor-semiconductor junctions~\cite{Camino}. Hereafter we will show how the fluctuating potentials in TLL affect this supercurrent shot noise. With use of the Green's functions defined by $\hat{g}^{>(<)}=\bigl( \hat{g}^{K} \pm (\hat{g}^{R}-\hat{g}^{A}) \bigr)/2$, the current-current correlation function can be written as~\cite{Averin2,Khlus}
%%%%%%%%%%%%%%%%%%%%%%%%%%%%%%%%%%%%%%%%%%%%%%%%%%%%%%%%%%%%%%%%%%%%%%%%%%%%%%%
%%%%%%%%%%%%%%%%%%%%%%%%%%%%%%%%%   kernel   %%%%%%%%%%%%%%%%%%%%%%%%%%%%%%%%%%
\begin{eqnarray}
\hspace{-2mm}
K(t,t+\tau)=-\frac{e^2}{8}{\rm Tr}
\bigl[
\hat{g}^{>}(t,t+\tau|\phi)\hat{\tau}_z \hat{g}^{<}(t+\tau,t|\phi)\hat{\tau}_z 
\nonumber \\
+\hat{g}^{<}(t,t+\tau|\phi)\hat{\tau}_z \hat{g}^{>}(t+\tau,t|\phi)\hat{\tau}_z
\bigr].
\label{ccc}
\end{eqnarray}
%%%%%%%%%%%%%%%%%%%%%%%%%%%%%%%%%%%%%%%%%%%%%%%%%%%%%%%%%%%%%%%%%%%%%%%%%%%%%%%
Here we focus on zero frequency spectral density of the current fluctuation $S(0)=\int d\tau /(2\pi) \overline{\langle K(t,t+\tau) \rangle _{\phi}}$. The bar over $K$ indicates the average over the time $t$. For simplicity, we disregard the Andreev reflections for $|\epsilon |>\Delta$ and the relaxations in the superconductors. 

Physically the $\theta$ fields play a similar role to the {\it measuring environment}, which is introduced to compute electron counting statistics~\cite{Levitov}. Hence the functional average of Eq.~(\ref{ccc}) over them gives the Gaussian statistics of the charge number in the 1D region. The resultant zero frequency spectral density is found to be
%%%%%%%%%%%%%%%%%%%%%%%%%%%%%%%%%%%%%%%%%%%%%%%%%%%%%%%%%%%%%%%%%%%%%%%%%%%%%%%
%%%%%%%%%%%%%%%%%%%%%%%%%%%%%  zero frequency   %%%%%%%%%%%%%%%%%%%%%%%%%%%%%%%
\begin{eqnarray}
\frac{S(0)}{S_0}&=&
{\rm Re}
\int d\epsilon d\epsilon'\sum_{m=0}^{\infty} 
\frac{P_{m}(\epsilon')}{\Delta}
\nonumber \\
&\times&
\prod_{l=1}^{m} 
\Bigl(
e^{-i\frac{\epsilon'}{v_F/ L}}
\gamma_R
\bigl(
\epsilon-l eV
\bigr)
\gamma_R^*
\bigl(
\epsilon+\epsilon'-l eV
\bigr)
\Bigr),
\nonumber \\
\label{supernoise}
\end{eqnarray}
%%%%%%%%%%%%%%%%%%%%%%%%%%%%%%%%%%%%%%%%%%%%%%%%%%%%%%%%%%%%%%%%%%%%%%%%%%%%%%%
%%%%%%%%%%%%%%%%%%%%%%%%%%%%%%%%%%%%%%%%%%%%%%%%%%%%%%%%%%%%%%%%%%%%%%%%%%%%%%%
with $S_0=e^2\Delta/(2\pi^2 \cosh^2(\Delta/2T))$. The function $P_m(\epsilon)$ describes the energy exchange between an electron and the internal environment. Within the lowest order of $(v_F/LD)$, it is given by
%%%%%%%%%%%%%%%%%%%%%%%%%%%%%%%%%%%%%%%%%%%%%%%%%%%%%%%%%%%%%%%%%%%%%%%%%%
\begin{eqnarray}
P_m (\epsilon)=\left\{
\begin{aligned}
\frac{1}{2}\delta (\epsilon)&&  (m=0) \\
C_{m-1}&+&2C_m+C_{m+1} ~~~~(m \not= 0)
\end{aligned}
\right.
\end{eqnarray}
%%%%%%%%%%%%%%%%%%%%%%%%%%%%%%%%%%%%%%%%%%%%%%%%%%%%%%%%%%%%%%%%%%%%%%%%%%
where $\delta (\epsilon)$ is Dirac's delta function, and
%%%%%%%%%%%%%%%%%%%%%%%%%%%%%%%%%%%%%%%%%%%%%%%%%%%%%%%%%%%%%%%%%%%%%%%%%%
\begin{eqnarray}
C_m \sim \frac{1}{2\pi D}(\frac{\pi T}{D})^{\beta_m-1}
\frac{\cosh (\frac{\epsilon}{2 T})}{\Gamma(\beta_m)}
\left|
\Gamma(\frac{\beta_m}{2}+i\frac{\epsilon}{2\pi T})
\right|^2.
\label{cm}
\end{eqnarray}
%%%%%%%%%%%%%%%%%%%%%%%%%%%%%%%%%%%%%%%%%%%%%%%%%%%%%%%%%%%%%%%%%%%%%%%%%%
Equation~(\ref{cm}) reminds us of the transition rate derived by Fermi's golden rule in Caldeira-Leggett model~\cite{Caldeira}. This shows that the internal fluctuations disturb the superconducting phase coherence. The exponent on the temperature is expressed by
%%%%%%%%%%%%%%%%%%%%%%%%%%%%%%%%%%%%%%%%%%%%%%%%%%%%%%%%%%%%%%%%%%%%%%%%%%
\begin{eqnarray}
\beta_m=\left\{
\begin{aligned}
\frac{m^2}{2} (K_{\rho}^{-1}-1)&&  (m; {\rm even}) \\
\frac{m^2}{2} (K_{\rho}^{-1}-1)&+&\frac{1}{2}(K_{\rho}-1) ~~~~(m; {\rm odd}). 
\end{aligned}
\right.
\end{eqnarray}
%%%%%%%%%%%%%%%%%%%%%%%%%%%%%%%%%%%%%%%%%%%%%%%%%%%%%%%%%%%%%%%%%%%%%%%%%%
The additional exponent in odd $m$ process is originated in the phase field $\alpha_{a,s}(\tau)=\tilde{\alpha}_{a,s}(0)-\tilde{\alpha}_{a,s}(\tau)$ with
%%%%%%%%%%%%%%%%%%%%%%%%%%%%%%%%%%%%%%%%%%%%%%%%%%%%%%%%%%%%%%%%%%%%%%%%%%
\begin{eqnarray}
\tilde{\alpha}_{a,s}(\tau)=\frac{1}{2}
\Bigl[
\theta_{a,s}(\frac{L}{2},\tau)-\theta_{-a,-s}(\frac{L}{2},\tau)
-\{ \frac{L}{2}\rightarrow -\frac{L}{2} \}
\Bigr]
.
\nonumber 
\end{eqnarray}
%%%%%%%%%%%%%%%%%%%%%%%%%%%%%%%%%%%%%%%%%%%%%%%%%%%%%%%%%%%%%%%%%%%%%%%%%%
This implies that the difference in exponents for even and odd $m$ owes to the interference between the states before and after the multiple Andreev reflections. In the processes with odd number of the Andreev reflections, an injected electron-like quasiparticle comes back as a hole-like quasiparticle with the fluctuating correlations shouldering. Such an interference does not occur for even $m$ case because an injected quasiparticle transmits into the other electrode. Besides, when the repulsive interactions are absent ($K_{\rho}=1$), $C_m=\delta(\epsilon)/4$ and the result in Ref.~\cite{Averin2} is rightly reproduced.
%%%%%%%%%%%%%%%%%%%%%%%%%%%%%%%%%%%%%%%%%%%%%%%%%%%%%%%%%%%%%%%%%%%%%%%%%%%%%%%
\begin{figure}[b]
\includegraphics[scale=0.58]{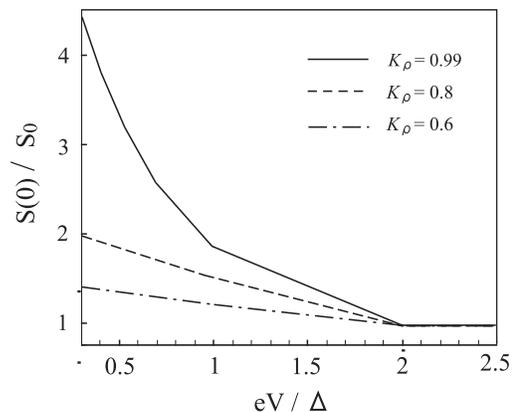}
\caption{
Zero frequency spectral densities are plotted as functions of $eV/\Delta$ for different $K_{\rho}s$. Here we set the parameters as $T=0.2 \Delta\sim 0.1D$ and $L\sim 1.2 v_F/\Delta$.
}
\label{noise}
\end{figure}
%%%%%%%%%%%%%%%%%%%%%%%%%%%%%%%%%%%%%%%%%%%%%%%%%%%%%%%%%%%%%%%%%%%%%%%%%%%%%%%

Figure~\ref{noise} illustrate the zero frequency spectral densities as functions of the bias voltage. One can see that the repulsive interactions slack the gradient of the shot noise at $eV < 2\Delta$. Moreover, in the low bias limit $eV \ll \Delta$, we can replace the summation in Eq.~(\ref{supernoise}) by the integration. This enables us to have the asymptotic behavior of the supercurrent fluctuation. Assuming that the low energy excitations ($\epsilon \ll \Delta$) predominantly influence the shot noise, the zero frequency spectral density above can be written approximately as
%%%%%%%%%%%%%%%%%%%%%%%%%%%%%%%%%%%%%%%%%%%%%%%%%%%%%%%%%%%%%%%%%%%%%%%%%%
\begin{eqnarray}
S(0) \sim S_0 
\Bigl[
1+\frac{\Delta}{eV} R
\Bigr],
\end{eqnarray}
%%%%%%%%%%%%%%%%%%%%%%%%%%%%%%%%%%%%%%%%%%%%%%%%%%%%%%%%%%%%%%%%%%%%%%%%%%
where
%%%%%%%%%%%%%%%%%%%%%%%%%%%%%%%%%%%%%%%%%%%%%%%%%%%%%%%%%%%%%%%%%%%%%%%%%%
\begin{eqnarray}
R=
\int d\epsilon'
\frac{\cos(\frac{\epsilon'}{v_F/L})
+\cos(\frac{\epsilon'}{v_F/L}+\frac{\pi \epsilon'}{eV})}{1-(\epsilon'/eV)^2}
P_{\frac{n_c}{3}}(\epsilon').
\label{shotreno}
\end{eqnarray}
%%%%%%%%%%%%%%%%%%%%%%%%%%%%%%%%%%%%%%%%%%%%%%%%%%%%%%%%%%%%%%%%%%%%%%%%%%
Here $n_c={\rm Int}[1+2\Delta/eV]$ is the number of possible Andreev reflections with ${\rm Int}[\cdots]$ denoting integer part. Although the factor $R$ somewhat overestimates the effect of TLL, it provides compendious scenario. In non-interacting limit, $S(0)$ is proportional to $n_c e$ which indicates the existence of large charge quanta. On the other hand, in the presence of the repulsive interactions, the coherence-origin excess noise exhibits a peak at some voltage and disappear as $eV\rightarrow 0$ owing to the considerably large power. Although it needs some corrections when the relaxations in the superconductors are taken into account~\cite{Averin2}, the peak structure is not qualitatively changed.

In summary, we have investigated the relation between low-lying fluctuations in TLL and AC Josephson effect. It was found that the microscopic excitations in 1D configuration can act as a kind of disturbance, and AC Josephson effect is essentially exposed to them. The repulsive interactions in TLL were found to inhibit Josephson oscillations and coherence-origin supercurrent noise. On the other hand, time-independent current is not influenced, which indicates the fluctuations act only on the phase difference. Recently, Titov {\it et al.} showed that the interactions renormalize the Andreev phase (not the Andreev reflection probability) with use of scaling approach~\cite{Titov}. Our result is consistent with theirs within quasiclassical approximation. 

In this work, we have restricted ourselves to the perfect transparency and the large capacitance limit. In tunneling limit, it is known that the proximity effect enhances the charge fluctuations~\cite{Takane}. Besides, in the regime in which charging energy becomes relevant, the effective action for $\theta$s has a mass term at the interfaces~\cite{Oshikawa}. In these situations, not only the average current but also the current noise will need the large corrections. We think that these are left for the interesting future problems.

We thank K. Kamide and Y. Terakawa for useful comments and discussions. This work is partly supported by a Grant for The 21st Century COE Program (Holistic Research and Education center for Physics of Self-organization Systems) at Waseda University from the Ministry of Education, Culture, Sports, Science and Technology of Japan.

\end{document}